\documentclass[journal=jacsat,manuscript=article]{achemso}
\usepackage[version=3]{mhchem} 
\usepackage[T1]{fontenc}       
\usepackage{subfig}
\usepackage{graphicx}
\usepackage{color}

\author{Benhui Yang}
\affiliation{Department of Physics and Astronomy and Center for
  Simulational Physics, University of Georgia, Athens, GA 30602}
\email{yang@physast.uga.edu}
\author{P. Zhang}
\affiliation{Department of Chemistry, Duke University, Durham, NC 27708}
\author{Chen Qu}
\affiliation{Department of Chemistry, Emory University, Atlanta, GA 30322}
\author{X. H. Wang}
\affiliation{Department of Chemistry, Emory University, Atlanta, GA 30322}
\author{P. C. Stancil}
\affiliation{Department of Physics and Astronomy and Center for
  Simulational Physics, University of Georgia, Athens, GA 30602}
\author{J. M. Bowman}
\affiliation{Department of Chemistry, Emory University, Atlanta, GA 30322}
\author{N. Balakrishnan}
\affiliation{Department of Chemistry, University of Nevada, Las Vegas,
  NV 89154}
\author{B. M. McLaughlin}
\affiliation{Centre for Theoretical Atomic, Molecular and Optical Physics (CTAMOP), 
  School of Mathematics and Physics, Queen's University Belfast, The David Bates Building, 
  7 College Park, Belfast BT7 1NN, United Kingdom}
\author{R. C. Forrey}
\affiliation{Department of Physics, Penn State University,
   Berks Campus, Reading, PA 19610}

\title {Full-dimensional Quantum Dynamics of SiO in Collision with H$_2$}


\begin{document}

\begin{abstract}
We report the first full-dimensional potential energy surface (PES) and quantum mechanical close-coupling
calculations for scattering of SiO due to H$_2$.  The full-dimensional interaction potential surface 
was computed using the explicitly correlated coupled-cluster (CCSD(T)-F12b) method 
and fitted using an invariant polynomial approach. 
Pure rotational quenching cross sections from initial states $v_1=0$, $j_1$=1-5 of SiO in collision 
with H$_2$ are calculated for collision energies between 1.0 and 5000 cm$^{-1}$. 
State-to-state rotational rate coefficients are calculated at temperatures between 5 and 1000 K.
The rotational rate coefficients of SiO with para-H$_2$ are compared with previous approximate results
which were obtained using SiO-He PESs or scaled from SiO-He rate coefficients. 
Rovibrational state-to-state and total quenching cross sections and rate coefficients
for initially excited SiO($v_1=1, j_1$=0 and 1) in collisions
with para-H$_2$($v_2=0,j_2=0$) and ortho-H$_2$($v_2=0,j_2=1$) were also obtained.
The application of the current collisional rate coefficients to astrophysics is briefly 
discussed.

\end{abstract}

\section{Introduction}

Molecular hydrogen is the most abundant species in most interstellar environments.
Collisional relaxation of rotationally or vibrationally excited
molecules by H$_2$ impact is therefore an important process in astrophysics, astrochemistry, and 
in many environments where non-equilibrium kinetics plays a dominant role.
In the interstellar medium (ISM), cooling processes are primarily associated with  
collisional thermal energy transfer between internal degrees of freedom followed by emission of radiation.
Collisional data for state-to-state vibrational and rotational quenching rate coefficients
are needed to accurately model the thermal balance and kinetics in the ISM.
Quantum mechanical scattering calculations are the primary source of these much needed 
collisional data \cite{flo11}.

A quantum close-coupling (CC) treatment has been developed \cite{tak65,pog02} 
for full-dimensional collisions involving two diatomic molecules.
With the recent development of the quantum scattering code TwoBC \cite{twobc}, 
which implements full angular momentum coupling,
it is now feasible to perform extensive rovibrational coupled-channel scattering calculations for 
diatom-diatom systems in full-dimensionality.
The first six-dimensional (6D) CC calculations of rovibrational collisions 
of H$_2$ with H$_2$ were presented recently \cite{que09,sam11,sam13}.  
Subsequently, full-dimensional CC computations were extended to the complex systems 
CO-H$_2$ \cite{co15,co16} and CN-H$_2$ \cite{cn16} 
on 6D PESs constructed from high-level ab initio electronic structure calculations. 
However, the full quantum close-coupling method 
is expensive due to the large number of basis states and coupled channels. 
Recently this difficulty has been alleviated to some extent by using 
the coupled-states (CS) approximation presented by Forrey and coworkers \cite{boh14,for15}.
The CS approximation has been successfully implemented in H$_2$+H$_2$ and CO+H$_2$ rovibrational scattering
calculations and achieved reasonable agreement with CC results \cite{boh14,for15}.
However, in this work we adopt the CC method. 

Interstellar silicon monoxide (SiO) was first detected by Wilson et al. \cite{wil71} through the line
emission of $j_1=3-2$ in Sgr B2. Recently, Fonfr\'{i}a et al. \cite{fon14} observed SiO 
transitions $(v_1 = 0, \ j_1 = 6-5)$ and  $(v_1 = 1,\  j_1 = 6-5)$ 
in the ground and first excited vibrational states towards the C-rich AGB star IRC +10216.
However, it was pointed out that fitting the emission of the molecular lines in the  warmest regions of 
the envelopes of AGB stars ($T_k \simeq 1000 - 3000$ K) is a challenging task due to the 
lack of collisional coefficients. 
Tercero et al. \cite{ter11} detected the $j_1=2-1$ and $j_1=4-3$ lines for excited $v_1=1$ state of SiO
in a survey towards Orion KL.
Using a radiative transfer code, they modeled the lines of the detected silicon-bearing species.
The SiO-H$_2$ rotational rate coefficients of Dayou and Balan\c{c}a \cite{day06},
derived from rigid-rotor approximation calculations on a SiO-He surface,
 were used in the modeling.
Prieto et al. \cite{pri17} detected for the first time toward IK Tau
rotational lines of SiO isotopologues in vibrationally excited states. 
Ag\'{u}ndez et al. \cite{agu12} observed $j_1 = 2-1$ through $j_1= 8-7$ rotational transitions of 
the $v_1 = 0$ state and three transitions of the $v_1 = 1$ state of SiO in the inner layers of IRC +10216.
In their radiative transfer modeling the rate coefficients of  Dayou and Balan\c{c}a \cite{day06} were adopted
for the first 20 rotational levels and for temperatures up to 300 K. 
However, for temperatures higher than 300 K and for rovibrational transitions,
the collision rate coefficients used for carbon monosulfide were adopted.
Justtanont et al. \cite{jus12} reported the observation of SiO lines
in the Herschel HIFI spectra of nine oxygen-rich AGB stars. These
included pure rotational lines ($j_1=14-13$ and $j_1=16-15$) and 
vibrationally excited $(v_1 = 1)$ rotational lines $j_1=13-12$, $j_1=15-14$, and $j_1=23-22$.
Matsuura et al. \cite{mat14} performed non-local thermodynamic equilibrium (NLTE)
 line radiative transfer calculations using
the SiO-H$_2$ collisional excitation rate coefficients which were obtained by scaling the
 SiO-He values of Dayou and Balan\c{c}a \cite{day06} with a factor of 1.38.  
 However, these radiative transfer calculations only included rotational transitions. 
Ignoring vibrationally excited transitions could potentially cause errors for molecular 
lines arising from high temperature gas, typically over 1000 K.  More recently, vibrational excitation 
calculations of SiO in collision with He has been reported.  
Using the vibrational close-coupling rotational infinite order sudden method, 
Balan\c{c}a and Dayou \cite{bal17} calculated vibrational de-excitation rate coefficients of
SiO from the first six vibrational levels. 

To our knowledge, no SiO-H$_2$ PES exists addressing the interaction between SiO and H$_2$.
Pure rotational (de)excitation rate coefficients for some selected rotational levels of SiO with
para-H$_2$ are available, but they were computed using an SiO-He PES and SiO-H$_2$ reduced mass. 
Turner et al. \cite{tur92}  calculated rotational excitation rate coefficients for SiO in collision 
with para-H$_2$($j_2$=0) using the coupled-states approximation method and a SiO-He PES obtained from 
an electron gas model \cite{bie81}. Dayou and Balan\c{c}a \cite{day06} constructed 
a 2D SiO-He potential energy surface based on highly correlated 
ab initio calculations.  This SiO-He PES was also used to compute rate coefficients for 
the rotational (de)excitation of SiO by collision with para-H$_2$ ($j_2$=0).
Even though these SiO-H$_2$ rate coefficients are approximate and not accurate, they are still used
in a variety of astrophysical modeling.  

Here we present the first full-dimensional PES for the SiO-H$_2$ complex. We also performed the first
scattering calculations for this system for rotational and vibrational inelastic processes in full dimension.
The paper is organized as follows. The theoretical methods are briefly described in Sec. II. 
The results are presented and discussed in Sec. III. Astrophysical applications are discussed in Section IV. 
Section V summarizes the results and presents an outlook on future work.

\section{Theoretical Methods}

In this section the theoretical methods used in the PES calculation and fit as well as
 rovibrational inelastic scattering calculation 
are briefly described. The reader is referred to Refs.~[\hspace*{-4px}\citenum{que09,co15,bra09}] 
for more details of the methodology. 

\subsection{Potential energy surface computations and fit}

\begin{figure}[h]
\advance\leftskip -0.0cm
\includegraphics[scale=0.65, angle=0]{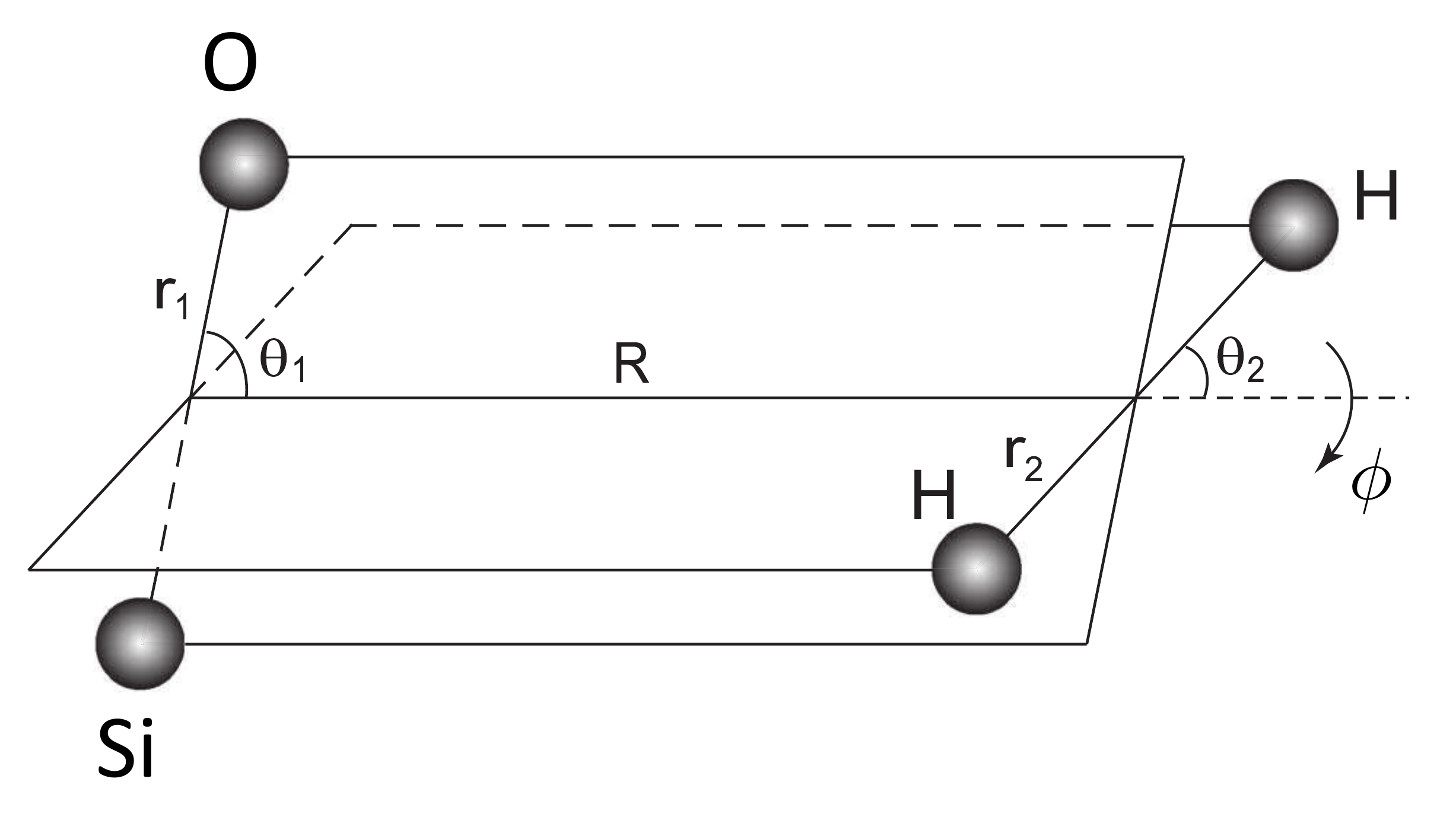}
\caption{The six-dimensional Jacobi coordinates for the SiO-H$_2$ system.}
\label{fig_jacobi}
\end{figure}

The 6D interaction potential of SiO-H$_2$  in the electronic ground state  
was calculated on a 6D grid using Jacobi coordinates 
($R, r_1, r_2, \theta_1, \theta_2, \phi$) as shown in Fig.~\ref{fig_jacobi}. 
The distance between the centers of mass of SiO and H$_2$ is denoted by $R$, 
while $r_1$ and $r_2$ refer to the bond lengths
describing the vibration of SiO and H$_2$, respectively. The angle $\theta_1$ ($\theta_2$) 
is the in-plane orientation angle
between $\vec{r}_1$ ($\vec{r}_2$) and $\vec{R}$, and $\phi$  the out-of-plane dihedral angle.
In the ab initio calculations, the intermolecular distance $R$
was chosen in the range of 4.0 - 21.0~$a_0$ and the bond distances are taken over the ranges
$2.5 \leq r_1 \leq 3.4$~$a_0$  and $1.0 \leq r_2 \leq 2.25$~$a_0$.  
Here $a_0$=0.529~\AA~ is the Bohr radius. The PES was computed
with the MOLPRO suite of computational chemistry codes \cite{molpro1,molpro2}.
 To produce an accurate 6D PES with proper symmetry
 the ab initio calculations  were performed over the angular coordinates 
$0 \leq \theta_1 \leq 360^\circ$ and $0 \leq \theta_2,\phi \leq 180^\circ$, 
where $\theta_1=\theta_2=0^\circ$ corresponds to the collinear configuration Si-O $-$ H-H. 

The ab initio electronic structure computations of potential energies were performed using the 
explicitly correlated coupled-cluster (CCSD(T)-F12b) method \cite{ccsdf12,densityfit}. 
All the calculations employed aug-cc-pVQZ (for H and O atoms) \cite{ken92} 
and aug-cc-pwCVQZ (for Si atom) orbital basis set \cite{pet02}, and the corresponding MP2FIT 
auxiliary bases \cite{wei02,hat05} for density fitting. The aug-cc-pV6Z-RI auxiliary 
bases (without k functions) \cite{emsl} were used for the resolutions of the identify and density-fitted Fock 
matrices for all orbital bases. Benchmark calculations at this CCSD(T)-F12 level were carried out 
on selected molecular configurations and results were compared with those from the conventional 
CCSD(T) method using aug-cc-pV5Z. The counter-poise (CP) \cite{cp} corrected interaction energy agrees 
closely with those derived from CCSD(T)/aug-cc-pV5Z. The interaction PES was corrected for basis 
set superposition error (BSSE) \cite{bsse}.  No scaled triples correction was used in 
our calculation.

The 6D SiO-H$_2$ interaction potential, referred to as VSiOH2, 
is a hybrid one that combines a fit to the full ab initio data set (denoted $V_{\text{I}}$) 
and a fit to the long-range data (denoted $V_{\text{II}}$) and is given by
\begin{equation}
V = (1-s) V_{\text{I}} + s V_{\text{II}},
\end{equation}
where $s$ is a switching function, defined as
\begin{equation}
s = \left\{
\begin{aligned}
& 0\ (R < R_i)\\
& 10 b^3 - 15 b^4 + 6 b^5 \ (R_i < R < R_f), \\
& 1\ (R > R_f)
\end{aligned}
\right.
\end{equation}
 $R_i=10.0$ $a_0$ and $R_f=12.0$ $a_0$, and $b = (R - R_i) / (R_f - R_i)$.\\

Both $V_{\text{I}}$ and $V_{\text{II}}$ have been fitted in 6D using an invariant polynomial 
method \cite{bra09,bow10} and are expanded in the form
\begin{equation}
V(y_1 \cdots y_6) = \sum_{n_1 \cdots n_6} C_{n_1 \cdots n_6} y_1^{n_1} y_6^{n_6} \left(
y_2^{n_2} y_3^{n_3} y_4^{n_4} y_5^{n_5} + y_2^{n_3} y_3^{n_2} y_4^{n_5} y_5^{n_4} \right)
\end{equation}
where $y_i = \exp(-d_i/p)$ are Morse-type variables, and $p$ is a user-specified parameter. 
For $V_{\text{I}}$ we used $p=3.0$ $a_0$, and for $V_{\text{II}}$ $p=9.5$ $a_0$.
 The internuclear distances $d_i$ between
two atoms are defined as $d_1 = d_\text{SiO}$, $d_2 = d_\text{SiH}$, $d_3 = d_\text{SiH}$,
$d_4 = d_\text{OH}$, $d_5 = d_\text{OH}$ and $d_6 = d_\text{HH}$.
The powers $n_1, \cdots, n_6$ satisfy $n_1 + \cdots + n_6 \leq 7$ and $n_2 + n_3 + n_4 + n_5 \neq 0$, 
that is, the maximum power of the polynomial is 7, and the interaction potential is guaranteed 
to go to zero when SiO and H$_2$ are separated to $R\to \infty$ for all $r_1$ and $r_2$.
The total number of linear coefficients $C_{n_1 \cdots n_6}$ is 882, and these coefficients were 
determined via linear least-squares fitting using our software MSA. \cite{xie10}
 The root mean square (RMS) fitting error in the long range
fit VII is 0.05 cm$^{-1}$,  for VI the RMS error is 2.61 cm$^{-1}$.
This hybrid approach greatly improves the behavior of the PES in the long range. 

\begin{figure}
\includegraphics[scale=0.7]{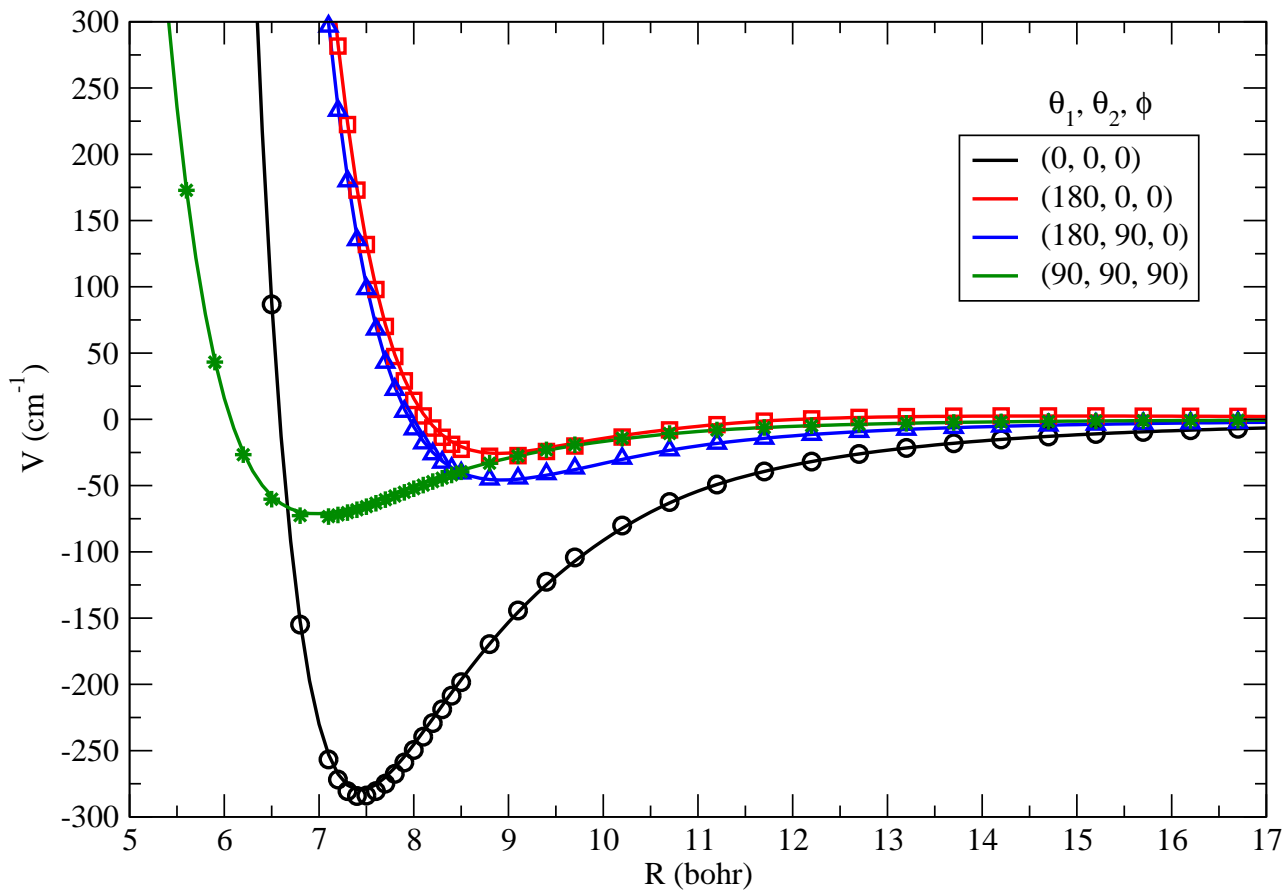}
\caption{The $R$ dependence of the SiO-H$_2$ PES, VSiOH2 for 
    ($\theta_1, \theta_2, \phi$)=
    (0$^{\circ}$, 0$^{\circ}$, 0$^{\circ}$),
    (180$^{\circ}$, 90$^{\circ}$, 0$^{\circ}$),
    (90$^{\circ}$, 90$^{\circ}$, 90$^{\circ}$),
   and (180$^{\circ}$, 0$^{\circ}$, 0$^{\circ}$). The bond lengths of SiO and H$_2$ are fixed
  at their equilibrium distances. Symbols are ab initio energy points.}
\label{vr_angle}
\end{figure}

\begin{figure}
\centering
 \parbox{5.0in} {
\includegraphics[scale=0.6]{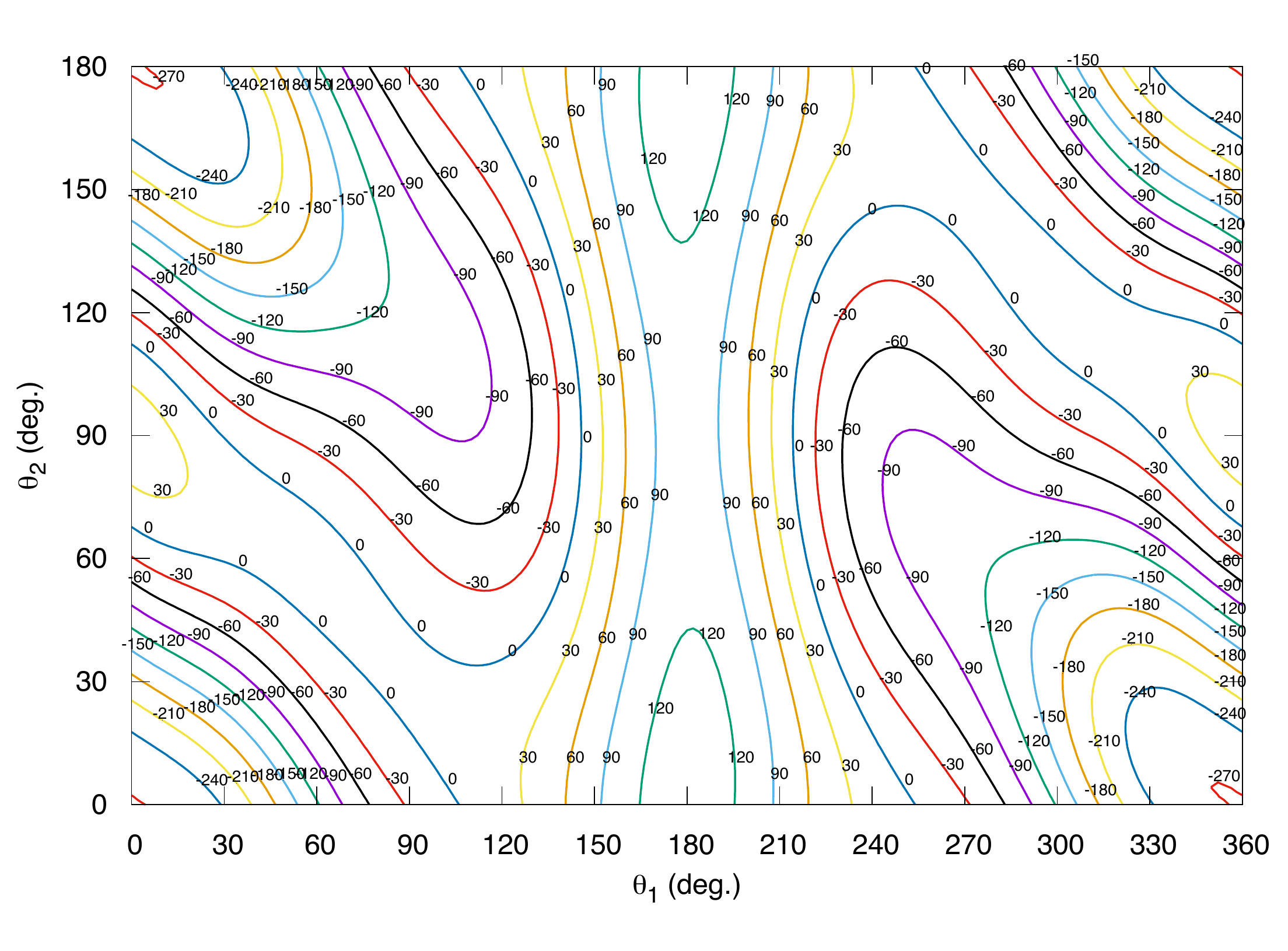}
}
\\
\begin{minipage}{5.0in} {
\includegraphics[scale=0.6]{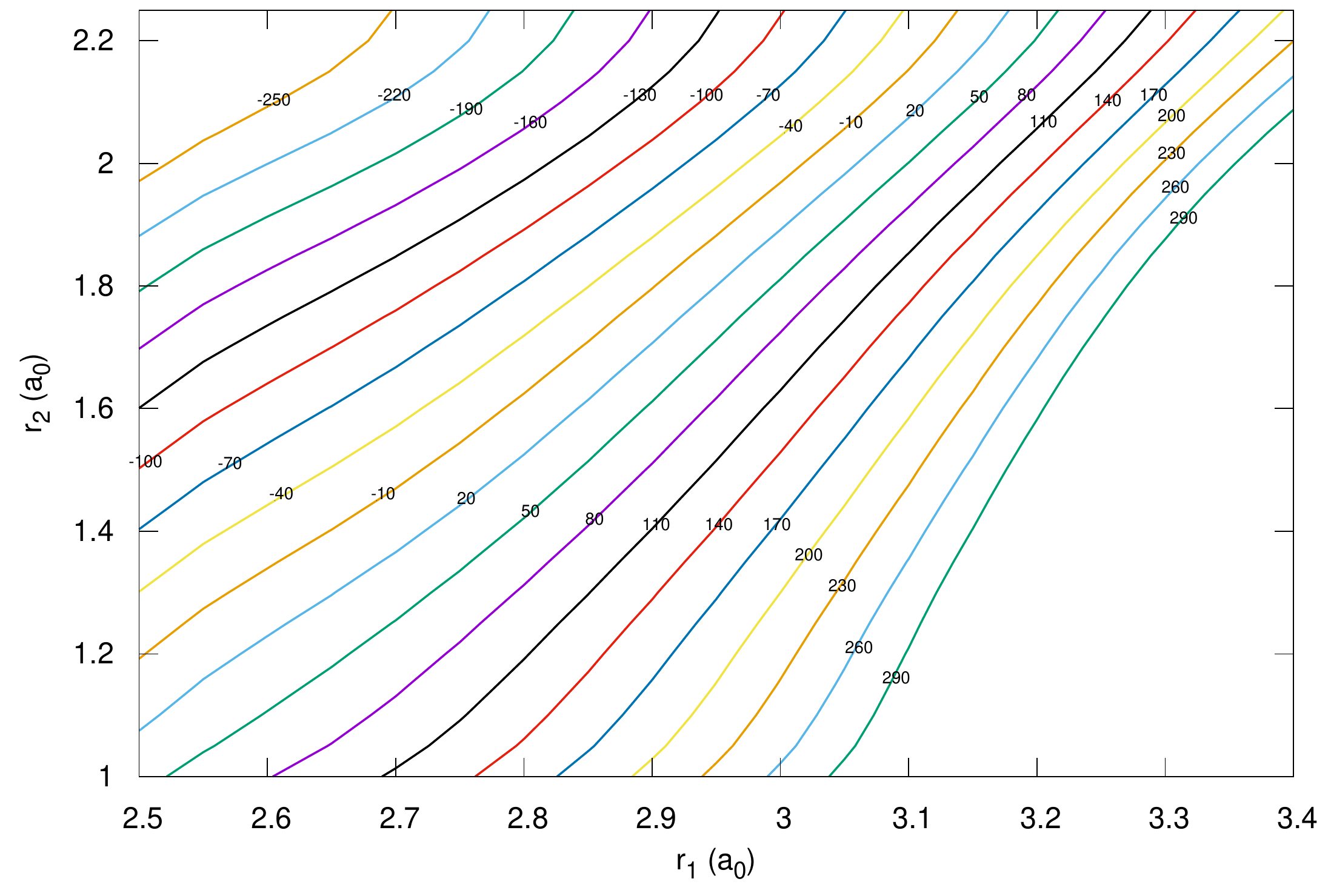}
}
\end{minipage}
\caption{Contour plots of the potential VSiOH2 as a function 
  of $\theta_1$ and $\theta_2$ (Upper panel) for
  $r_1$=2.8530 $a_0$, $r_2$=1.4011 $a_0$, $R$=7.5 $a_0$, and $\phi$=0$^{\circ}$;
  and of $r_1$ and $r_2$ (Lower panel) for 
   $R$=6.5 $a_0$ and $\theta_1$ = $\theta_2$ = $\phi$=0$^{\circ}$.
  Additional PES plots can be found in the Supporting Information.
 }
\label{contour}
\end{figure}

Fig.~\ref{vr_angle} shows the $R$ dependence of 6D PES 
for $(\theta_1, \theta_2, \phi)$= (0$^{\circ}$, 0$^{\circ}$, 0$^{\circ}$),
(180$^{\circ}$, 0$^{\circ}$, 0$^{\circ}$), (180$^{\circ}$, 90$^{\circ}$, 0$^{\circ}$),
and (90$^{\circ}$, 90$^{\circ}$, 90$^{\circ}$). Symbols are energy points from ab initio 
calculations. The good agreement between ab initio energy points and the fitted PES confirms the
accuracy of our fitted PES. 
In Fig.~\ref{contour} (upper panel) we show a two-dimensional contour plot of the VSiOH2 PES in 
$\theta_1$, $\theta_2$ space for fixed values of $r_1$=2.8530 $a_0$, $r_2$=1.4011 $a_0$, 
$R=7.5$ $a_0$, and $\phi=0^{\circ}$.  To show the dependence of the PES on $r_1$ and $r_2$,
a two-dimensional contour plot in $r_1$, $r_2$ space for fixed values of 
$R=7.5$ $a_0$, and $\theta_1=\theta_2=\phi=0^{\circ}$ is also displayed in the lower panel of
Fig.~\ref{contour}.
The coordinates for the global minimum of -279.5 cm$^{-1}$ for the fitted potential
 and -284.2 cm$^{-1}$ for the
ab initio data correspond to  $(R, r_1, r_2, \theta_1, \theta_2, \phi)$=
(7.4~$a_0$, 2.8530~$a_0$, 1.4011~$a_0$, 0$^{\circ}$, 0$^{\circ}$, 0$^{\circ}$)

\section*{Scattering Theory and Computational Details}

The quantum CC formalism for diatom-diatom scattering including
vibrational motion has been fully developed \cite{tak65,gre75,ale77,zar74} 
and applied to a number of full-dimensional scattering studies \cite{sam11,co15,cn16}.
To facilitate the scattering calculations, 
the interaction  potential between SiO and H$_2$, $V(\vec{r}_1,\vec{r}_2,\vec{R})$, 
which vanishes when SiO and H$_2$ are far apart, can be written as, 

\begin{equation}
V(R,r_1,r_2,\theta_1,\theta_2,\phi) = \sum^{}_{\lambda_1\lambda_2\lambda_{12}} 
A_{\lambda_1\lambda_2\lambda_{12}}(r_1,r_2,R)  Y_{\lambda_1\lambda_2\lambda_{12}}
(\hat{r}_1,\hat{r}_2,\hat{R}) ,
\label{v-lbd}
\end{equation}
with the bi-spherical harmonic function expressed as,
\begin{eqnarray}
\lefteqn{Y_{ \lambda_1\lambda_2\lambda_{12}}(\hat{r}_1,\hat{r}_2,\hat{R}) =
\sum^{}_{m_{\lambda_1}m_{\lambda_2}m_{\lambda_{12}}} \big\langle  \lambda_1 m_{\lambda_1} \lambda_2 m_{\lambda_2} \big| 
 \lambda_{12} m_{\lambda_{12}} \big\rangle} \nonumber \\
 & & \ \ \ \ \ \ \ \ \  \times Y_{\lambda_1 m_{\lambda_1}}(\hat{r}_1) Y_{\lambda_2 m_{\lambda_2}}(\hat{r}_2) 
 Y^*_{\lambda_{12} m_{\lambda_{12}}}(\hat{R}),
\end{eqnarray}
where $ 0 \leq \lambda_1 \leq 8$, $0 \leq \lambda_2 \leq 4$.  
Only even values of $\lambda_2$  contribute  due to the homonuclear symmetry of H$_2$.
 
A combined molecular state (CMS) \cite{que09} notation, $(v_1j_1 v_2 j_2)$ was applied to describe a 
combination of rovibrational states for SiO $(v_1 j_1)$ and H$_2$ $(v_2 j_2)$.  
The quantum numbers $j$ and $v$ denote the rotational and vibrational energy levels.
 The state-to-state rovibrational cross section can be expressed  
as a function of the collision energy $E_c$,
\begin{eqnarray}
\lefteqn{\sigma_{v_1j_1v_2j_2 \to v'_1j'_1v'_2j'_2}(E_c) = \frac{\pi}{(2j_1+1)(2j_2+1)k^2}} \nonumber  \\
&& \ \ \ \ \ \ \ \ \ \ \ \  \times \sum_{j_{12}j'_{12}ll'J\varepsilon_I}^{} (2J+1) 
|\delta_{v_1j_1v_2j_2l,v_1'j_1'v_2'j_2'l'} 
 - S^{J\varepsilon_I}_{v_1j_1v_2j_2l,v_1'j'_1v_2'j_2'l'}(E_c)|^{2},
\label{s2scross}
\end{eqnarray}
where ($v_1j_1v_2j_2$) and ($v_1^{\prime} j_1^{\prime}v_2^{\prime}j_2^{\prime}$) denote  
the initial and final CMSs, respectively. The wave vector 
$k=\sqrt{2 \mu E_c/\hbar^2}$, and $S$ is the scattering matrix. The quantum number $l$ denotes the
orbital angular momentum, the total angular momentum $\vec{J}$ is given by
 $\vec{J}=\vec{l}+\vec{j}_{12}$, where $\vec{j}_{12} = \vec{j}_1 + \vec{j}_2$.

The total quenching cross section of SiO from initial state $(v_1 j_1 v_2 j_2) \to
(v_1^{\prime}; v_2^{\prime} j_2^{\prime})$ was obtained
by summing the state-to-state quenching cross sections over the final rotational state of
 $j_1^{\prime}$ of SiO in vibrational state $v_1^{\prime}$,
\begin{equation}
\sigma_{v_1 j_1 v_2 j_2 \to v_1^{\prime}; v_2^{\prime} j_2^{\prime}}(E_c)
    = \sum_{j_1^{\prime}} \sigma_{v_1j_1v_2j_2 \to v'_1j'_1v'_2j'_2}(E_c).
\label{totalcross}
\end{equation}

The state-to-state rate coefficients at a temperature $T$
can be obtained by thermally averaging
the corresponding integral cross sections over a Maxwellian kinetic energy distribution,

\begin{equation}
k_{v_1j_1v_2j_2\rightarrow v_1^{\prime} j_1^{\prime}v_2^{\prime}j_2^{\prime}}(T)
= \left (\frac{8}{\pi \mu \beta} \right )^{1/2}\beta^2\int^{\infty}_0 E_c
\sigma_{v_1j_1v_2j_2\rightarrow v_1^{\prime} j_1^{\prime}v_2^{\prime}j_2^{\prime}}(E_c)
\exp(-\beta E_c)dE_c,
\label{eq_rate}
\end{equation}
where $\mu$ is the reduced mass of the SiO-H$_2$ complex,
$\beta=(k_{\rm B}T)^{-1}$, and $k_{\rm B}$ is Boltzmann's constant.

Full-dimensional  rovibrational scattering calculations were carried out 
using the TwoBC code \cite{twobc}
in which the CC equations propagated for each value of $R$  from 4 to 21.0~$a_0$  
using the log-derivative matrix propagation method of Johnson \cite{joh73}.
The number of Gauss-Hermite quadrature points $N_{r_1}$, $N_{r_2}$; the number of 
Gauss-Legendre quadrature points in $\theta_1$ and $\theta_2$,
$N_{\theta_1}$, $N_{\theta_2}$; and the number of Chebyshev quadrature points in 
$\phi$, $N_{\phi}$ adopted to   
project out the  expansion coefficients of the PES are listed in Table~\ref{parameters}.
For the monomer potentials, we used the results of Barton et al. \cite{bar13} for SiO 
and Schwenke \cite{sch88} for H$_2$.

\begin{table}
\begin{center}
\caption{Parameters used in the scattering calculations.} 
\vskip 0.5cm
\label{parameters}
\begin{tabular}{c @{\hspace{0.1cm}}  c @{\hspace{0.3cm}} c  @{\hspace{0.3cm}} c @{\hspace{0.3cm}} 
    c @{\hspace{0.3cm}} c @{\hspace{0.3cm}} c @{\hspace{0.6cm}} c } 

\hline \hline  \\ [0.3ex]
   & Basis set
   & $N_{\theta_1}$($N_{\theta_2}$) & $N_{\phi}$ & $N_{r_1}$($N_{r_2})$
    & $\lambda_1$ &  $\lambda_2$  &  \\ [0.3ex]  
\hline \hline \\ [0.5ex]
 6D Rotation & & & & & &   \\
  \hline \\ [0.3ex]
para-H$_2$-SiO & $j_1=30$, $j_2=2 $ & 12 & 8 & 18  & 8 & 4 & (16, 30, 80, 230)\textsuperscript{\emph{b}} \\ [1ex]
ortho-H$_2$-SiO & $j_1=30$, $j_2=3$ & 12 & 8 & 18  & 8 & 4 &(18, 32, 82, 232)\textsuperscript{\emph{b}} \\ 
 \hline  \\ [0.3ex]
 6D Rovibration  & & & & & &    \\  \hline \\ [0.3ex]
 para-H$_2$-SiO   & [(0,35;1,20)(0,2)]\textsuperscript{\emph{a}} & 12   & 8 & 18  & 8  & 4 & 
   (16, 30, 80, 230)\textsuperscript{\emph{b}}   \\   [1ex]
 ortho-H$_2$-SiO   & [(0,35;1,20)(0,3)]\textsuperscript{\emph{a}} & 12  & 8 & 18  & 8 & 4 & 
   (18, 32, 82, 232)\textsuperscript{\emph{b}} \\   [0.3ex]
\hline \hline
\end{tabular}
\end{center}

\textsuperscript{\emph{a}} Basis set 
[($v_1=0, j_{v_1=0}^{\text{max}}$; $v_1=1, j_{v_1=1}^{\text{max}}$)($v_2=0, j_{v_2=0}^{\text{max}}$)]
is presented by the maximum rotational quantum number $j_{v_1}^{\text{max}}$ and $j_{v_2}^{\text{max}}$ 
included in each relevant vibrational level $v_1$ and $v_2$ for SiO and H$_2$, respectively. \\ 
\vskip 0.3cm
\textsuperscript{\emph{b}} Maximum partial waves $J(J_{E_1}, J_{E_2}, J_{E_3}, J_{E_4})$ 
used in scattering calculations for collision energies $E_1$=10, $E_2$=100, 
$E_3$=1000, and $E_4$=5000 cm$^{-1}$,
 respectively. 
\end{table}

\section{Results and Discussion}

\subsection{Pure rotational excitation}

\begin{figure}
\includegraphics[scale=0.7]{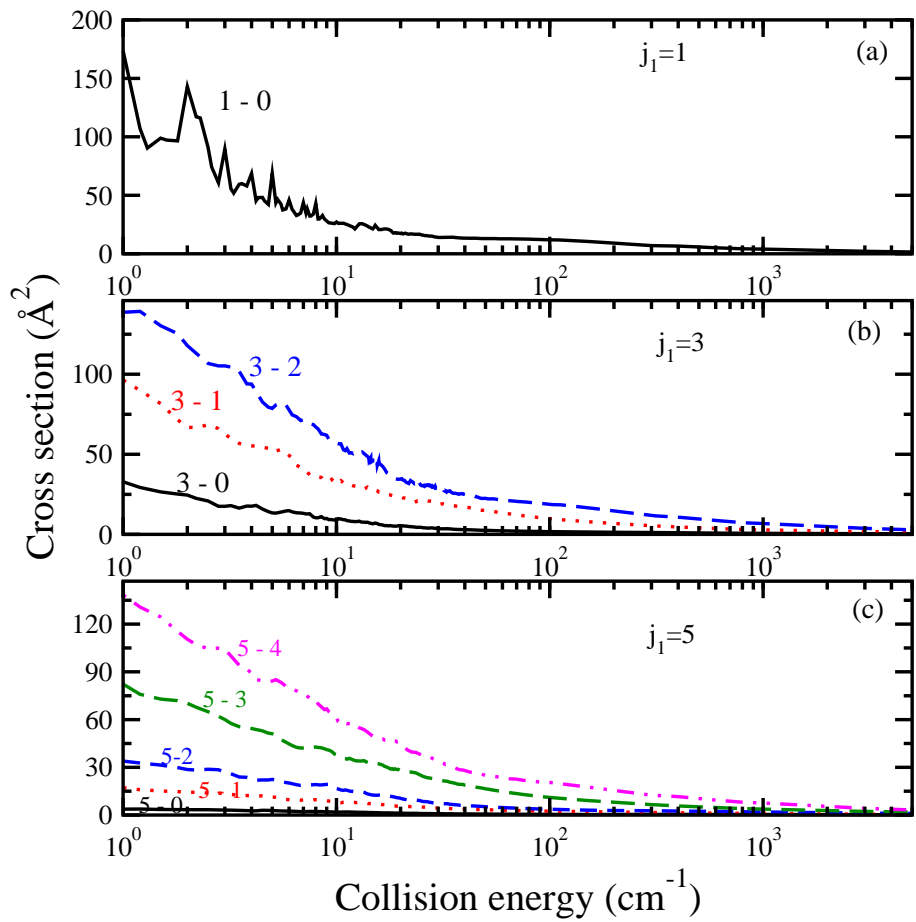} 
\caption{ 
Rotational state-to-state de-excitation cross sections for SiO($j_1$) + H$_2$($j_2$=1) $\rightarrow$
SiO ($j_1^{\prime}$) + H$_2$($j_2^{\prime}$=1), $j_1$=1, 3, and 5, $j_1^{\prime} < j_1$.
}
\label{cso135}
\end{figure}

The rotational excitation and deexcitation cross sections of SiO by H$_2$ 
have been calculated for the process, 
SiO($v_1$=0, $j_1$)+H$_2$($v_2$=0, $j_2$) $\rightarrow$ SiO($v_1^{\prime}=0$, $j_1^{\prime}$)
 + H$_2$($v_2^{\prime}=0$, $j_2^{\prime}$).
The calculations were carried out in full-dimension with the VSiOH2 PES and TwoBC code, with
both SiO and H$_2$ being in their ground vibrational states $v_1=v_2=0$.
The initial rotational states of SiO are $j_1=0 - 5$, with collision energy ranging from 
1 to 5000 cm$^{-1}$. 
The cross sections with respect to partial wave summation are converged to within $\sim$5\%,
which is ensured by varying the maximum number of partial waves considered (see Tabel 1).
The basis sets used in the scattering calculations are given in Table~\ref{parameters}.
Fig.~\ref{cso135} displays the state-to-state quenching 
cross section of SiO with para-H$_2$ ($j_2$=0) from initial $j_1$=1, 3, and 5. 
It is seen that for all initial $j_1$ states, the state-to-state quenching cross sections are dominated 
by $\Delta j_1=j_1^{\prime}-j_1=-1$ transitions.  The cross sections generally increase 
with increasing $j_1^{\prime}$ with the smallest cross sections corresponding
to transitions to $j_1^{\prime}$=0 (largest $\Delta j_1$),  
following a typical exponential energy-gap law behavior.  All of the 
cross sections display some resonances, particularly at collision energies below 20 cm$^{-1}$
due to quasibound states supported by the attractive part of the interaction potential.
It is also found that the cross sections are comparable for ortho- 
and para-H$_2$ colliders. \footnote{Figures for other initial states, as well as for ortho-H$_2$ 
colliders can be found in the Supporting Information.}

\begin{figure}
\includegraphics[scale=0.7]{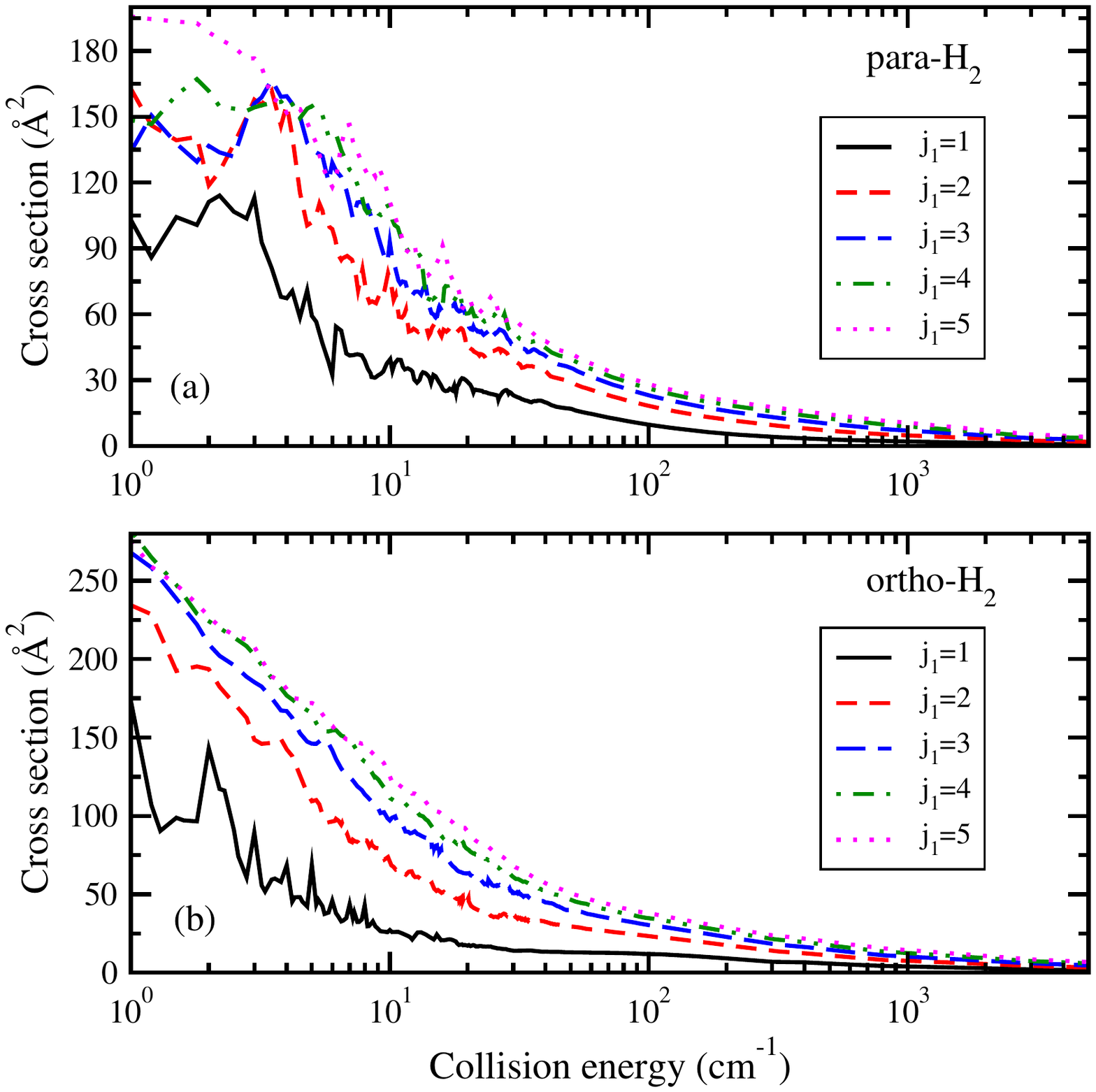} 
\caption{ 
 Total rotational de-excitation cross section for SiO from initial rotational states 
 $j_1$= 1 - 5 in collisions with (a) para-H$_2$($j_2$=0) and (b) ortho-H$_2$($j_2$=1). 
}
\label{totcsrot}
\end{figure}

Figs.~\ref{totcsrot} (a) and (b) illustrate total rotational quenching cross section 
of SiO with para- and ortho-H$_2$, respectively.  
It is seen that except for low collision energies below $\sim 5$ cm$^{-1}$, 
the total quenching cross sections increase with increasing initial $j_1$. 
The resonances present in the state-to-state cross sections can also be observed
in the total quenching cross sections.


\begin{figure}
\includegraphics[scale=0.7]{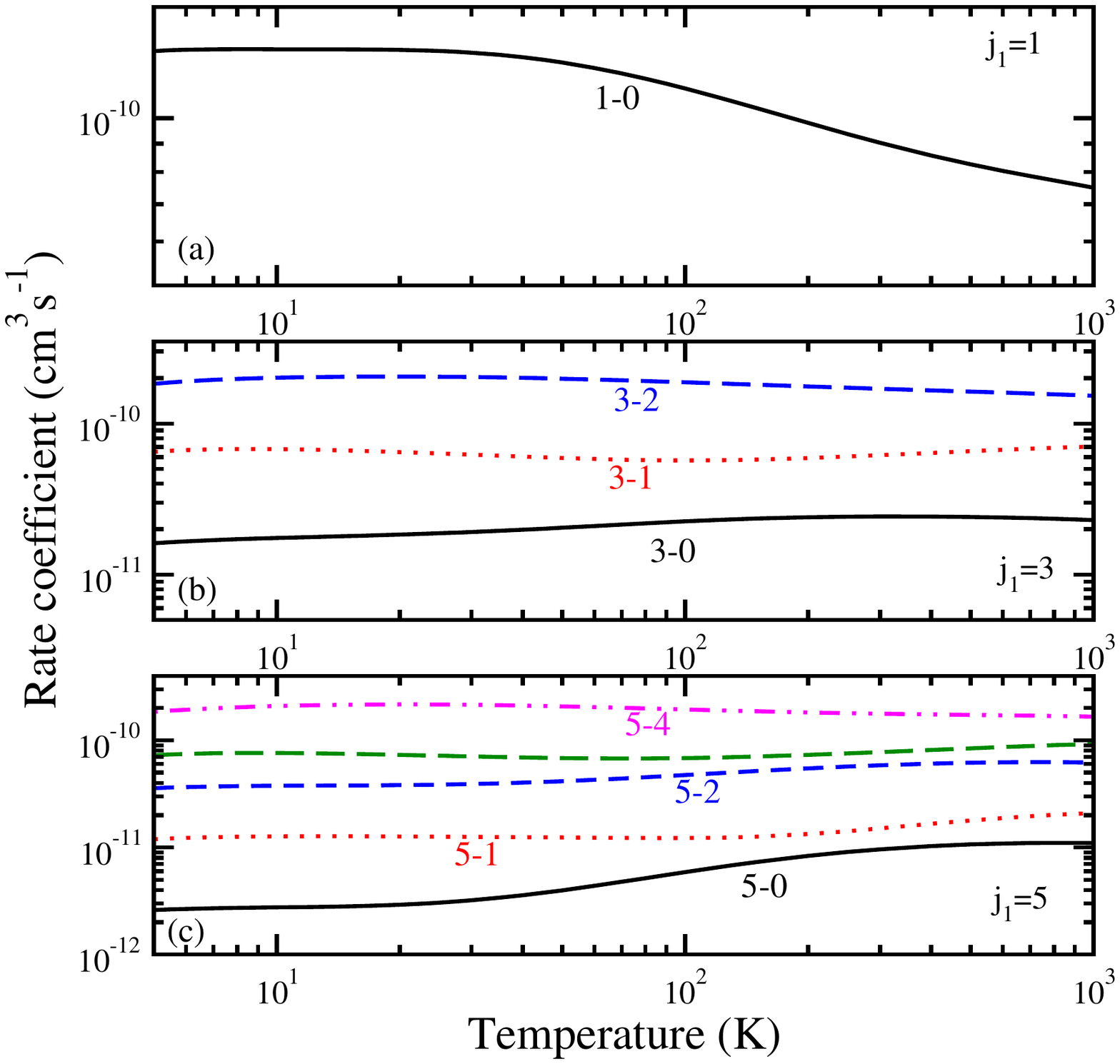} 
\caption{ 
Rotational state-to-state de-excitation rate coefficients for SiO($j_1$) + H$_2$($j_2$=0) $\rightarrow$
SiO ($j_1^{\prime}$) + H$_2$($j_2^{\prime}$=0), $j_1$=1, 3, and 5, $j_1^{\prime} < j_1$.
}
\label{ratep135}
\end{figure}

State-to-state rate coefficients for temperatures
between 5 and 1000 K were computed for SiO initial rotational states $j_1$=1 - 5.
As an example, in Fig.~\ref{ratep135} the state-to-state quenching rate coefficients 
from initial rotational states $j_1$=1, 3, and 5 are displayed
for SiO in collision with para-H$_2$ ($j_2$=0).  
Fig.~\ref{ratep135} shows that for both colliders 
the rate coefficients decrease with increasing
$|\Delta j_1|=|j_1^{\prime}-j_1|$ with $\Delta j_1=-1$ being the dominant transitions.
For all the selected initial states, the rate coefficients are generally flat except for 
some undulations due to resonances in the cross sections mostly for 
 $\Delta j_1=-1$ transitions.

\begin{figure}
\includegraphics[scale=0.7]{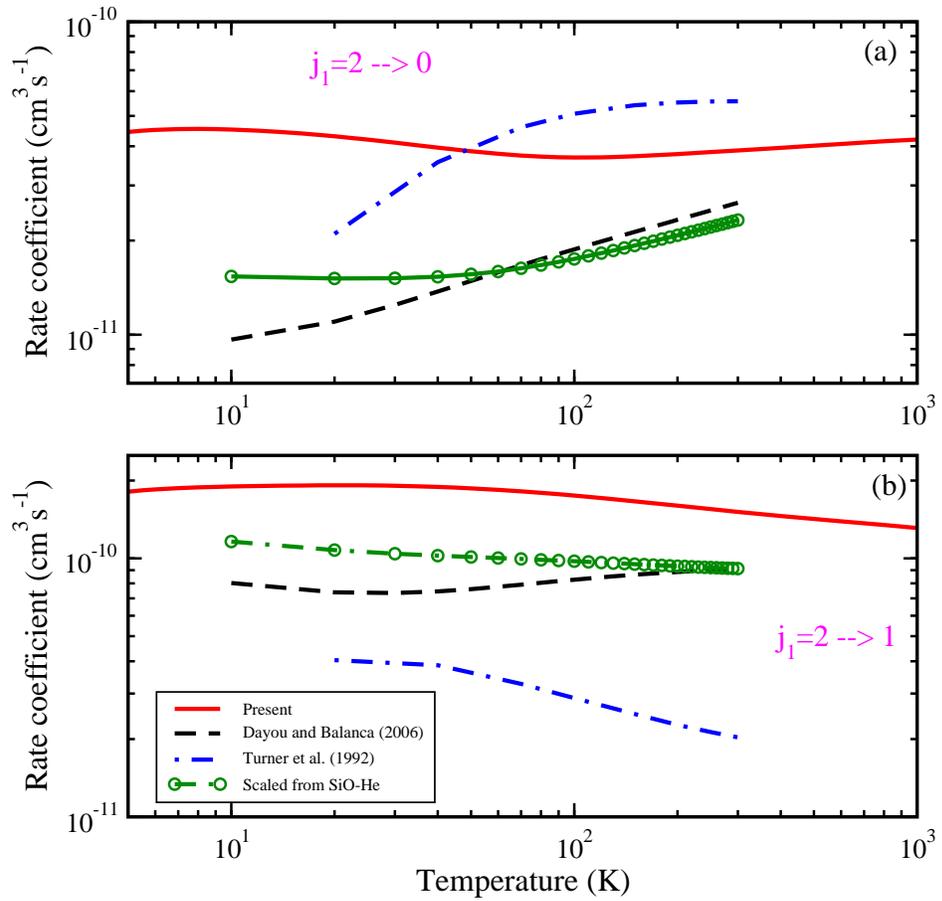} 
\caption{ 
Comparison of current state-to-state rotational quenching rate coefficients with previous 
approximate results from Refs.~[\hspace*{-4px}\citenum{tur92,day06}] and obtained from reduced mass scaling 
of SiO. Transitions are  from $j_1$=2 to $j_1^{\prime}$=0 and 1, and the collider is para-H$_2$.
}
\label{pdtsj2}
\end{figure}

\begin{figure}
\includegraphics[scale=0.7]{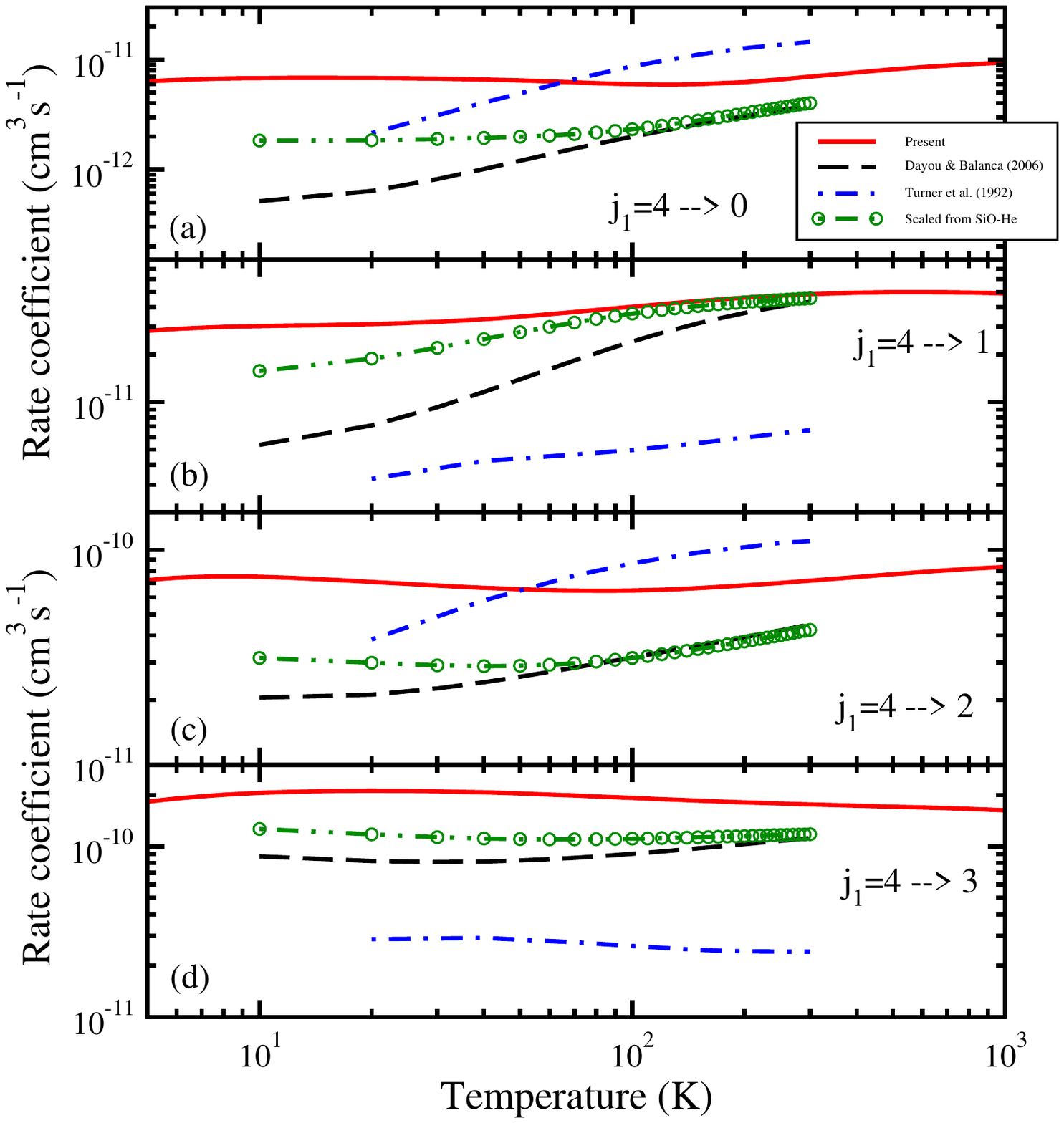} 
\caption{ 
Same as Fig.\ref{pdtsj2}, except that the rotational transitions
of SiO are from $j_1$=4 to $j_1^{\prime}$=0, 1, 2 and 3.
}
\label{pdtsj4}
\end{figure}

To the best of our knowledge, there have been no published experimental cross sections or
 rate coefficients available
for rotational transitions of SiO by collisions with H$_2$.
Theoretical studies are also very limited. For SiO in collision with para-H$_2$ ($j_2$=0),
pure rotational (de)excitation rate coefficients for some selected rotational levels are available,
but they were computed using SiO-He PESs.
Turner et al. \cite{tur92} calculated rotational excitation rate coefficients for SiO in collision
 with para-H$_2$($j$=0) using the coupled-states approximation and a SiO-He PES obtained from
 an electron gas model \cite{bie81}.
Dayou and Balan\c{c}a \cite{day06} constructed a 2D SiO-He potential energy surface.
This SiO-He PES was also used to compute rate coefficients for
the rotational (de)excitation of SiO by collision with para-H$_2$ ($j_2$=0).
Due to the assumed similarity of para-H$_2$ and He, the rate coefficient of SiO-para-H$_2$ ($j_2=0$)
may also be estimated from the corresponding results for collision with He using the reduced mass scaling
factor of 1.4 \cite{sch05}.  As examples, we compare  
in Figs.~\ref{pdtsj2} and \ref{pdtsj4} our accurate rotational rate coefficients with 
the approximate SiO-H$_2$ rate coefficients obtained using SiO-He PESs \cite{tur92,day06} and mass scaling.  
One can see that there are significant differences between the current state-to-state
 rate coefficients and
the approximate results. Our results are generally larger than the approximate ones, 
except for the deexcitation transitions
$j_1=2 \to 0$, $4 \to 0$, and $4 \to 2$ of Turner et al. which become the largest 
at temperatures above $\sim 60$ K.  
It is also worth noting that the global minimum of the 2D SiO-He PES of Dayou and Balan\c{c}a \cite{day06}
is -26.596 cm$^{-1}$ and is much shallower than the global well depth of VSiOH2 PES.
Therefore, the large discrepancies indicate that the use of an SiO-He PES or the simple 
mass scaling method are not suitable to estimate
the rate coefficients of SiO with para-H$_2$, 
while a more accurate scaling method may be adopted \cite{wal14}.

\subsection{Rovibrational quenching}

The main focus of this work is actually on full-dimensional calculations of 
the state-to-state cross sections for SiO rovibrational transitions from $v_1$=1, 
SiO($v_1$=1, $j_1$)+H$_2$($v_2$=0, $j_2$) $\rightarrow$ SiO($v_1^{\prime}=0$, $j_1^{\prime}$)
 + H$_2$($v_2^{\prime}=0$, $j_2^{\prime}$). 
For collision energies ranging from 1 to 5000 cm$^{-1}$, 
the basis sets used in the scattering calculations are listed in Table~\ref{parameters}.
The state-to-state cross sections are summed over SiO final rotational levels $v_1^{\prime}=0, j_1'$ 
to yield total vibrational quenching cross sections.
In the present calculations, $j_1$=0 and 1, $j_1^{\prime}$=0, 1, 2, $\cdots$, 35.  $j_2$=0 for 
para-H$_2$ and 1 for ortho-H$_2$.
Only the rotational transitions of H$_2$ are considered, the vibration of H$_2$
is fixed in the ground state $v_2=v_2^{\prime}=0$.

\begin{figure}
\includegraphics[scale=0.6]{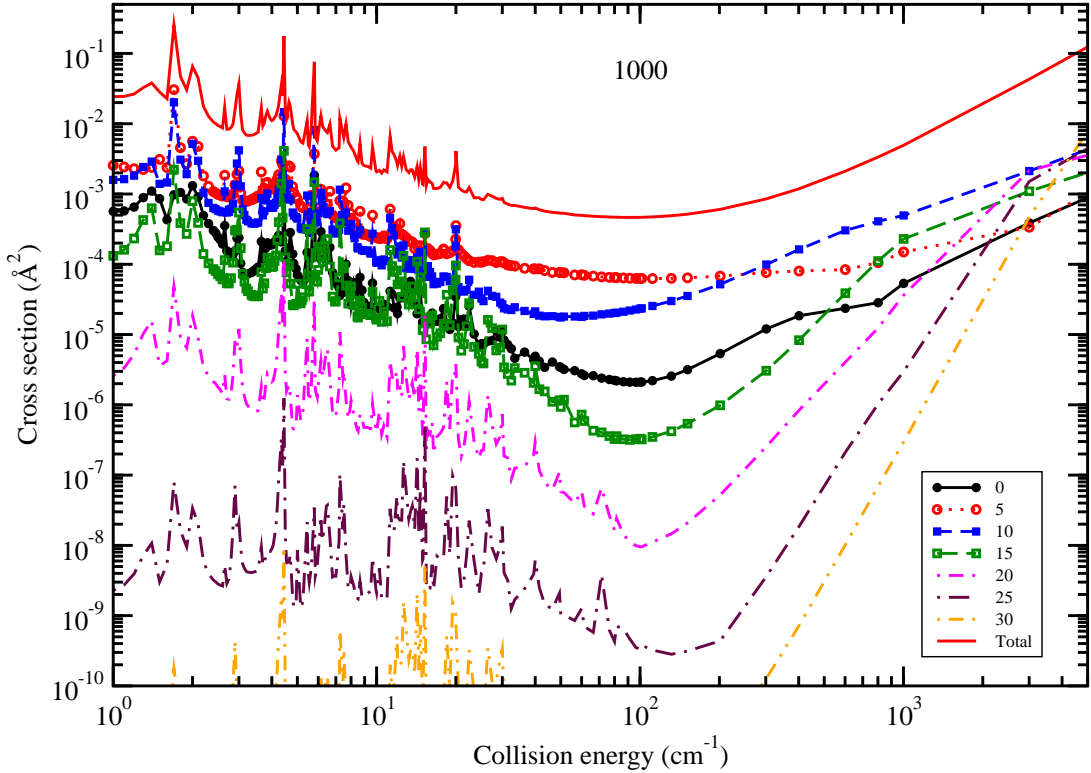} 
\caption{ 
State-to-state and total vibrational quenching cross section for SiO in collisions with para-H$_2$
 with H$_2$ elasticity ($j_2=0\rightarrow 0$). The initial state is (1000) and 
final states are $v_1^{\prime}=0$, $j_1^{\prime}$=0, 5, 10, $\cdots$, 30.
}
\label{s2s1000}
\end{figure}

For illustration, 
Fig.~\ref{s2s1000} presents the state-to-state and total cross sections for quenching 
from initial CMS (1000) into different final SiO rotational levels in $v_1^{\prime}=0$,
where we show only $j_1^{\prime}$=0, 5, $\cdots$, 30,  $j_2=0 \to j_2^\prime$=0. 
It can be seen that for  para-H$_2$ collider
 the cross sections present a large number of resonances
at energies between 1.0 and 200 cm$^{-1}$, the resonances extend to larger collision energy with
increasing $j_1^{\prime}$. In general, when the collision energies exceed the van der Waals well depth,
the state-to-state and 
total quenching cross sections increase with increasing collision energy.  
As shown in Fig.~\ref{s2s1000}, for the case of para-H$_2$, the cross section to $j_1^{\prime}$=5 dominates
at energies below $\sim 200$ cm$^{-1}$, then the cross section to $j_1^{\prime}$=10 becomes the largest for
collision energy above 200 cm$^{-1}$. At $E_c$=5000 cm$^{-1}$, the cross section to $j_1^{\prime}$=0 
is the smallest, while the cross section to  $j_1^{\prime}$=30 becomes the largest.  
 Cross sections for $j_1'$=20, 25, and 30 are several orders of magnitude 
smaller than the other transitions shown in Fig.~\ref{s2s1000} for energies below 500 cm$^{-1}$.
The cross sections for ortho-H$_2$ show similar behavior (see Supporting Information). 

\begin{figure}
\includegraphics[scale=0.7]{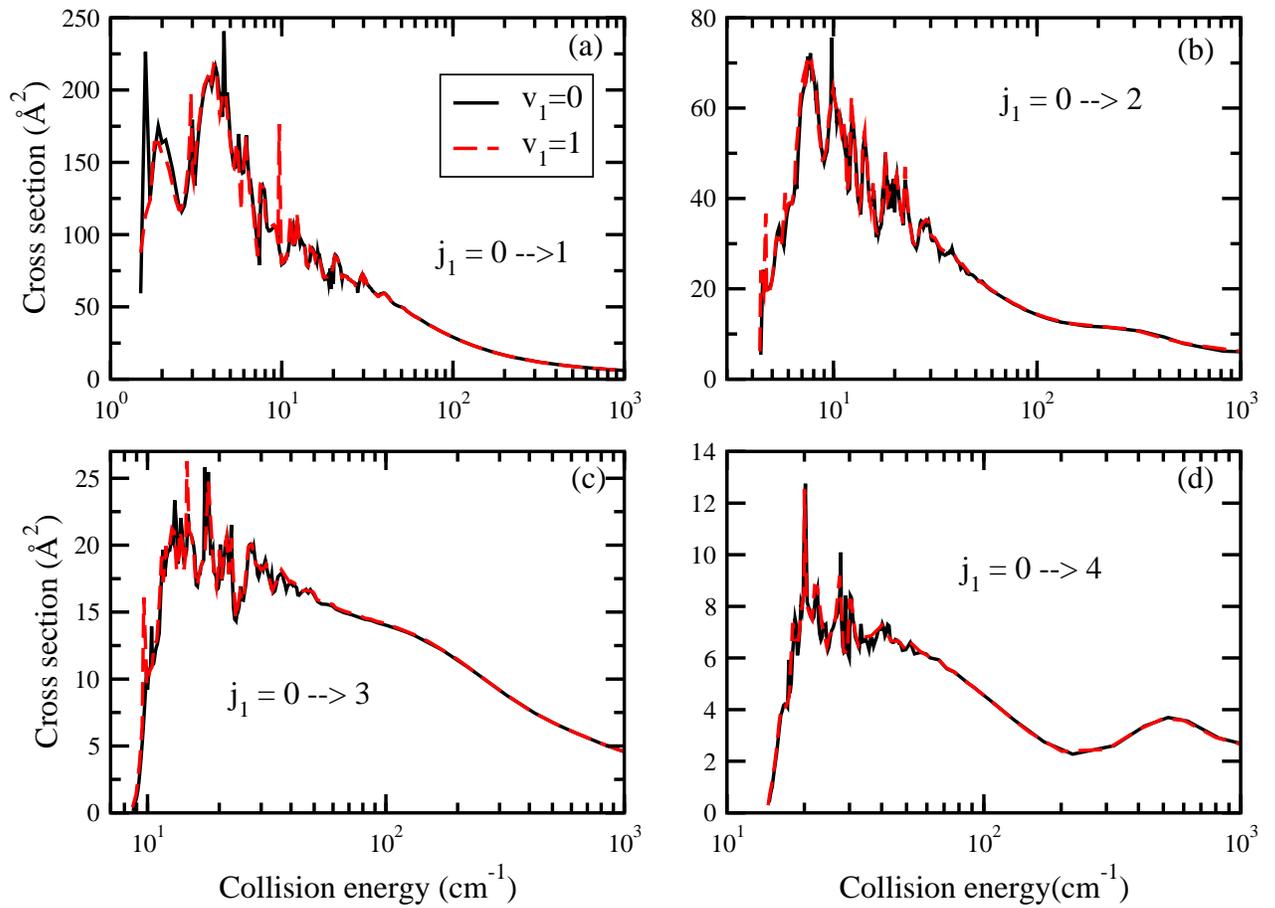} 
\caption{ 
Rotational excitation cross section from $j_1$=0 to $j_1^{\prime}$=1, 2, 3, and 4
 of SiO in collisions with para-H$_2$.
For each transition, the comparison is made between $v_1=0$ and $v_1=1$.
}
\label{purerotv10}
\end{figure}

To show the effect of vibrational excitation on pure rotational cross sections, 
in Fig.\ref{purerotv10} we compare cross section in $v_1$=0 and $v_1$=1 states.
The rotational excitations are from $j_1$=0 to $j_1^{\prime}$=1, 2, 3, and 4.  
It can be seen that the pure rotational cross sections are nearly identical for $v_1=0$ and 1,
except that cross sections for $v_1=0$ show strong resonances at low energy.
In other words, vibrational excitation has little effect on pure rotational transitions, which
might be expected due to the harmonic behavior of the monomer SiO potential.

\begin{figure}
\includegraphics[scale=0.6]{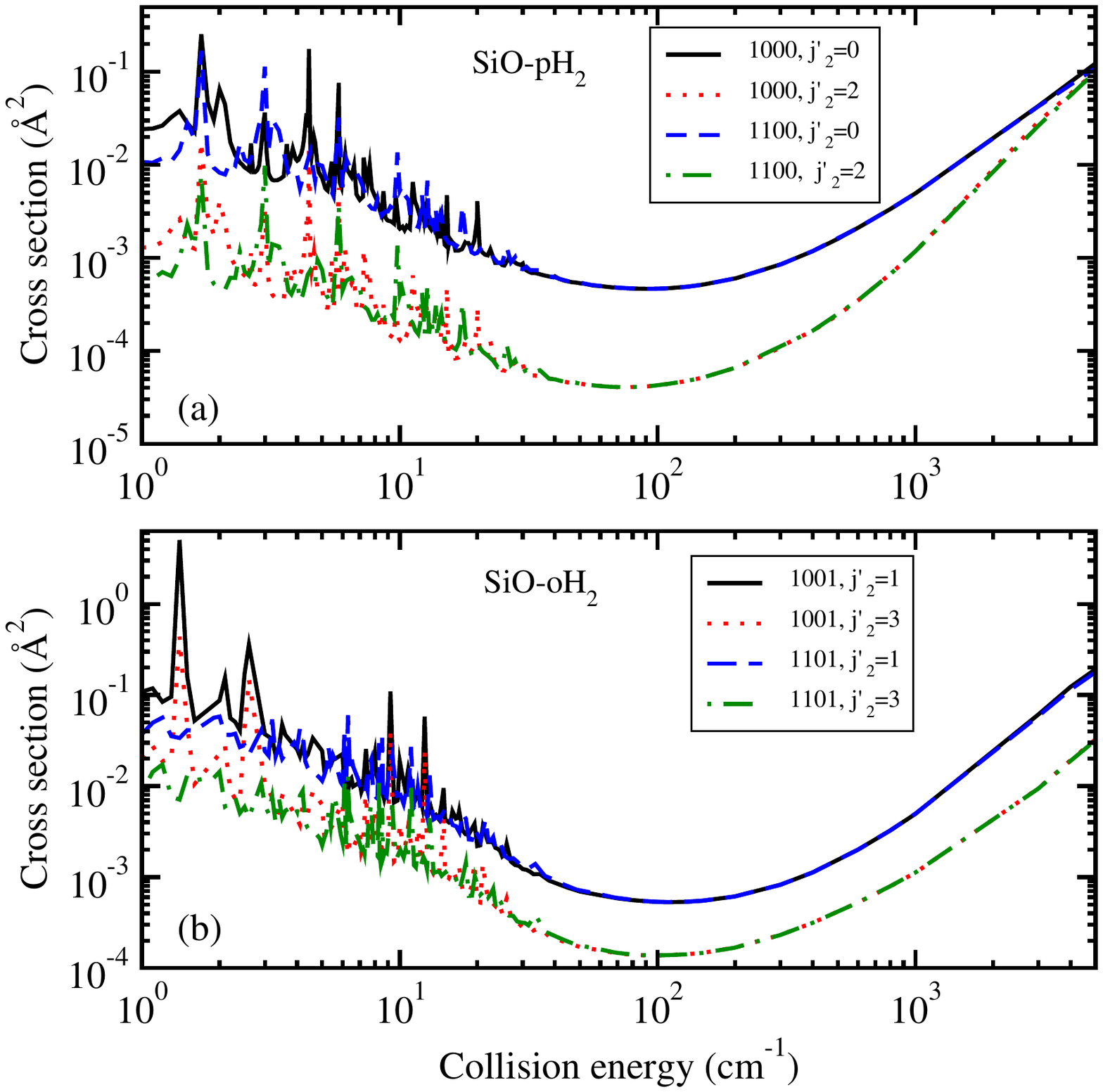} 
\caption{ 
  Total vibrational quenching cross sections for SiO in collisions with H$_2$. 
 (a) Initial states (1000) and (1100), final $j_2^{\prime}$=0 and 2;
 (b) initial states (1001) and (1101), $j_2^{\prime}$=1 and 3. 
}
\label{totcs_o_p}
\end{figure}

The vibrational quenching cross sections from initial states (1100) and (1101) are also calculated,
in which the excited $j_1$=1 levels are considered.
Fig.~\ref{totcs_o_p} (a) illustrates the energy dependence of the total $v_1=1 \to v_1^{\prime}=0$ quenching 
cross section with para-H$_2$ from CMSs (1000) and (1100) for H$_2$ elastic 
($j_2=0 \to j_2^{\prime}=0$) and inelastic ($j_2=0 \to j_2^{\prime}=2$) transitions.
The total quenching cross sections from (1000) and (1100) show some differences 
for energies less than 40 cm$^{-1}$,
in particular due to different resonance behavior. However, the total quenching cross sections are nearly
identical for energies above 40 cm$^{-1}$, which is likely due to the small energy difference 
between $j_1$=0 and $j_1$=1.  
The cross section corresponding to inelastic H$_2$ transition $j_2=0 \to 2$
are about one order of magnitude smaller than that for elastic H$_2$ transition for energies 
below 100 cm$^{-1}$. 
From Fig.~\ref{totcs_o_p} (b) it can be observed that, for the collision with ortho-H$_2$
the total vibrational quenching cross sections from initial CMSs (1001) and (1101) show
similar behavior.  The cross section with an elastic H$_2$ transition dominates,  
while the cross sections for $j_2=1 \to 3$ are about two times smaller than the H$_2$ rotation preserving
$j_2=1 \to 1$ transition. Furthermore, the difference becomes larger 
with increasing energy, therefore the cross section for the $j_2=1 \to 3$ transition  
can be neglected at high collision energies.

The state-to-state and total vibrational quenching rate coefficients are obtained by thermally
averaging corresponding cross sections over a Maxwellian distribution of kinetic 
energy. There are no published theoretical or experimental 
rate coefficients available. 

\begin{figure}[h]
\advance\leftskip -1.0cm
\includegraphics[scale=0.75, angle=0]{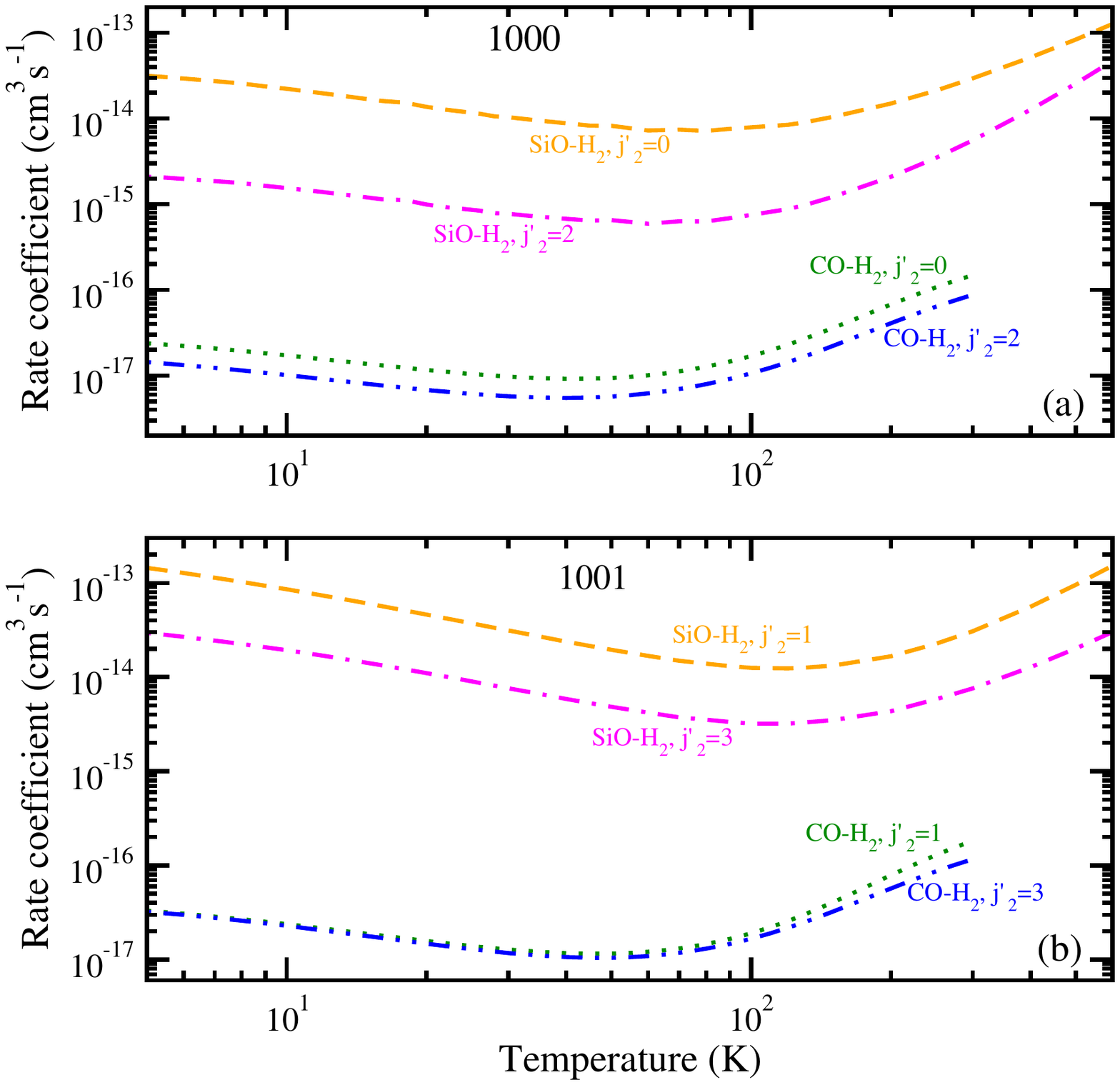}
\caption{
 Total rate coefficients for the vibrational quenching of SiO compared to the same
 transitions for CO from Ref.~[\hspace*{-5px}\citenum{co15}]. 
(a) from (1000) to $v_1^{\prime}$=0 + para-H$_2$($v_2^{\prime}$=0, $j_2^\prime$=0, 2).
(b) from (1001) to $v_1^{\prime}$=0 + ortho-H$_2$($v_2^{\prime}$=0, $j_2^\prime$=1, 3).
}
\label{rate_tot}
\end{figure}

As an illustration, we present the total rate coefficients for vibrational quenching 
from (1000) to ($v_1^{\prime}=0$) in collisions with  para-H$_2$($j_2^\prime$=0 and 2)
and from (1001) to ($v_1^{\prime}=0$) with ortho-H$_2$($j_2^\prime$=1 and 3)
as displayed in Fig.~\ref{rate_tot}. 
It can be seen that the total quenching rate coefficients of SiO with H$_2$ show
similar trends to that presented for the total quenching cross sections in Fig.~\ref{totcs_o_p}.
Fig.~\ref{rate_tot} (a) shows that for SiO with para-H$_2$,
the total vibrational quenching rate coefficients for $\Delta j_2=0$
are nearly an order of magnitude larger than the results for  $\Delta j_2=2$,
between 5 and $\sim$100 K the rate coefficients generally decrease weakly with
increasing temperature.  For temperatures above $\sim$ 100 K, the rate coefficients generally
increase with increasing temperature. As shown in Fig.~\ref{rate_tot} (b),
the trends for ortho-H$_2$ are very similar to those noted for para-H$_2$ collisions. 
The total vibrational quenching rate coefficients for $\Delta j_2=0$
are nearly three times larger than the results for  $\Delta j_2=2$.

In Fig.~\ref{rate_tot} we also compare to results for CO-H$_2$ (Ref.\hspace*{-4px}\citenum{co15})
where it is shown that for the same transitions, CO-H$_2$ rate coefficients are typically 
$\sim$2-3 orders of magnitude smaller than those of SiO-H$_2$.  The large magnitude of SiO-H$_2$ 
rate coefficients are likely related to the fact that the SiO-H$_2$ PES is more anisotropic.
Further, the overall magnitude of the cross sections increases with increasing global well depth:
93.1(CO-H$_2$) and 279.5 cm$^{-1}$ (SiO-H$_2$).
The vast difference between the CO and SiO rate coefficients suggests that scaling arguments based
on chemical similarities (see Ref.~[\hspace*{-4px}\citenum{tak11}]) should be used with caution.

\section{Astrophysical Applications}

Emission from interstellar molecular species are used to probe the physical and chemical conditions
of the ISM. In particular, observations of rotational and vibrational transitions
can provide information on elemental abundances, gas temperature, radiation field,
and other local parameters. Since
molecular hydrogen is the most abundant species in most cool astrophysical environments,
it is usually the dominant collider for molecular emission. While SiO is one of the
dominant silicon-containing molecules, elemental silicon is highly depleted onto dust
grains in most environments. Nevertheless, SiO emission, including maser lines, have
been observed.
SiO was first detected by observing the $j_1=3-2$ transition near 130.2 GHz with
the 11 m Kitt Peak telescope toward Sgr B2(OH) \cite{wil71}.
de Vicente et al. \cite{dev16} presented observational results of $^{28}$SiO $v_1=$0, $j_1=1-0$ line
emission from 28 evolved stars.
Wang et al. \cite{wan14} detected SiO $j_1=2-1 \ (v_1=3)$ mega-maser at 85.038 GHz 
near the center of Seyfert 2 galaxy NGC 1068 with the IRAM 30-m telescope.
Using the Mopra Telescope of the Australia Telescope National Facility,
Indermuehle and McIntosh \cite{ind14}  measured SiO spectra from the $v_1=1, \ j_1=1-0$
and the $v_1=1, \ j_1=2-1$ transitions.
Desmurs et al. \cite{des14} reported observations of the $v_1 = 1$, $v_1 = 2$,
and $v_1 = 3$ $j_1 = 1-0$ maser
transitions of SiO in several asymptotic giant branch (AGB) stars using very long baseline interferometry.
Two new SiO maser sources, in the high-mass star-forming regions
G19.61-0.23 and G75.78+0.34, were detected by Cho et al. \cite{cho16},
though SiO masers are rare in star-forming regions.

Since silicon is depleted onto grains, it is not detected in protoplanetary disks,
though CO is. It is proposed \cite{tan14}, however, that in the outflows of young stellar objects (YSOs), that
the grains are disrupted, allowing for the formation of SiO. Observations of SiO may therefore
be used to distinguish between the outflow and disk  of an YSO. In particular, SiO vibrational
emission may trace the wide-angle wind outflow.

However, modeling such spectra requires
collisional rate coefficients due to collision by the most abundant species H$_2$, H, and He.
Since, it is difficult to measure collisional rate coefficients, astrophysical modeling
mainly relies on theoretical results.
In previous studies discussed above, approximate SiO-para-H$_2$ rate coefficients, which were obtained
using an SiO-He PES were adopted, will lead to significant modeling uncertainty. The current
 collisional rate coefficients, therefore, will be critical to
advance astrophysical modeling of SiO observations.

\section{Summary}
Full-dimensional quantum close-coupling calculations of rotational and vibrational 
quenching of SiO in collision with para-H$_2$ ($j_1$=0) and ortho-H$_2$ ($j_1$=1) 
have been performed for the first time.
The considered initial rotational states of SiO are $j_1$=1-5 in $v_1=0$ and 
 $j_1$=0 and 1 in $v_1=1$ for collisions with 
 para-H$_2$ ($j_2=0, \ 2$) and ortho-H$_2$ ($j_2=1, \ 3$). 
The scattering calculations were carried out on
a 6D SiO-H$_2$ interaction potential surface computed using high-level ab initio 
theory and fitted with an invariant polynomial approach.  
Pure rotational quenching rate coefficients of SiO in collision with para-H$_2$
were compared with previous approximate results obtained using SiO-He potentials 
or mass-scaling methods. 
State-to-state and total quenching cross sections from the SiO 
vibrational state $v_1=1$ show resonance structures at intermediate energies for
both para-H$_2$ and ortho-H$_2$. 
The state-to-state rate coefficients for both rotational and vibrational quenching were computed
for temperatures ranging from 5 to 1000 K.  Calculations of quenching from higher excited 
rotational and vibrational states of SiO are in progress.
The current calculations together with large-scale coupled-states (CS) approximation \cite{for15} 
results will be essential in the construction of a database of SiO rotational and vibrational quenching 
rate coefficients urgently needed for astrophysical modeling \cite{agu12,jus12,mat14,zha17}.

\begin{suppinfo}

\begin{itemize}

\item
Fig.~S1. Anisotropy of the VSiOH2 PES for $R$= 6.0, 6.5, 7.0, and 8.0 $a_0$.

\item
Fig.~S2. Rotational state-to-state de-excitation cross sections for SiO with para-H$_2$, $j_1$=1, 3, and 5.

\item
Fig.~S3. Rotational state-to-state de-excitation cross sections for SiO with para-H$_2$, $j_1$= 2 and 4.

\item
Fig.~S4. Rotational state-to-state de-excitation cross sections for SiO with ortho-H$_2$, $j_1$= 2 and 4.

\item
Fig.~S5. Rotational state-to-state de-excitation rate coefficients for SiO with ortho-H$_2$, $j_1$=1, 3, 
and 5.

\item
Fig.~S6. State-to-state and total vibrational quenching cross section for SiO in collisions with ortho-H$_2$.

\item
Fig.~S7. Comparison of the distributions of final rotational levels in $v_1^{\prime}$ = 0
 quenching from CMSs (1000) and (1001) at $E_c$=1.0, 10.0, and 100.0 cm$^{-1}$.

\item
Fig.~S8. State-to-state and total vibrational quenching rate coefficients for SiO in collisions with
  ortho-H$_2$.

\item
A Fortran subroutine for generating the 6D VSiOH2 PES is available to download.

\end{itemize}

\end{suppinfo}

\section{AUTHOR INFORMATION}

\subsection{Corresponding Author}
*E-mail: yang@physast.uga.edu 

\subsection{Notes}
The authors declare no competing financial interest.

\section{ACKNOWLEDGMENTS}

 Work at UGA and Emory was supported by NASA grant NNX16AF09G,
 at UNLV by NSF Grant No. PHY-1505557, and at Penn State by NSF Grant No. PHY-1503615.
 B. Mclaughlin acknowledges the University of Georgia for travel funding  
 and Queens University Belfast for the award of a research fellowship.
 This study was supported in part by resources and technical expertise from the
 UGA Georgia Advanced Computing Resource Center (GACRC),  
 and UNLV National Supercomputing Institute \& Dedicated Research Network.
 We thank  Shan-Ho Tsai (GACRC),
 Jeff Deroshia (UGA Department of Physics and Astronomy),
 and Ron Young (UNLV) for computational assistance. 
 Portions of the potential surface calculations were performed at the National Energy Research
 Scientific Computing Center (NERSC) in Berkeley, CA, USA  and at the High Performance Computing
 Center Stuttgart (HLRS) of the University of Stuttgart, Stuttgart, Germany, where grants of time 
are gratefully acknowledged.

\end{document}